\documentclass[12pt,peerreview]{IEEEtran}

\IEEEtitleabstractindextext

\usepackage{subfig}
\usepackage{graphics}
\usepackage{graphicx}
\usepackage{color}

\usepackage{times}

\newenvironment{sciabstract}{%
\begin{quote} \bf}
{\end{quote}}

\renewcommand\refname{References and Notes}

\newcounter{lastnote}

\title{MAC Protocols for IEEE 802.11ax: \\ Avoiding Collisions on Dense Networks\footnote{{\bf Notice:} This work has been submitted to the IEEE for possible publication. Copyright may be transferred without notice, after which this version may no longer be accessible.}}

\author
{{\bf Rafael A. da Silva}$^{1}$ and {\bf Michele Nogueira}$^{1,2}$\\

\normalsize{$^{1}$Department of Informatics, Federal University of Paran\'a, Curitiba, PR, Brazil}\\
\normalsize{$^{2}$Electrical and Computer Engineering, Carnegie Mellon University, Pittsburgh, PA, USA}\\
\normalsize{E-mails: rasilva@inf.ufpr.br and michele@inf.ufpr.br}
}

\date{}

\begin{document} 


\baselineskip24pt


\maketitle 


\begin{sciabstract}


Wireless networks have become the main form of Internet access. Statistics show that the global mobile Internet penetration should exceed 70\% until 2019. Wi-Fi is an important player in this change. Founded on IEEE 802.11, this technology has a crucial impact in how we share broadband access both in domestic and corporate networks.  However, recent works have indicated performance issues in Wi-Fi networks, mainly when they have been deployed without planning and under high user density. Hence, different collision avoidance techniques and Medium Access Control protocols have been designed in order to improve Wi-Fi performance. Analyzing the collision problem, this work strengthens the claims found in the literature about the low Wi-Fi performance under dense scenarios. Then, in particular, this article overviews the MAC protocols used in the IEEE 802.11 standard and discusses solutions to mitigate collisions. Finally, it contributes presenting future trends in MAC protocols. This assists in foreseeing expected improvements for the next generation of Wi-Fi devices.
  
\end{sciabstract}


\section{Introduction}





Statistics show that wireless networks have become the main form of Internet access with an expected global mobile Internet penetration exceeding 70\% until 2019~\cite{ict2015statistics}. Wi-Fi is an important player in this change. Founded on IEEE 802.11 standard, this technology has a crucial impact in how we share broadband access both in domestic and corporate networks. However, over wireless medium, signals can suffer interference and disturbances from different sources. Further, network devices have to share the communication channel that, if not well managed, can produce unintelligible signals by concurrent transmissions on the same frequency.

Medium Access Control (MAC) protocols have been proposed to provide efficiency in the use of the communication channel. Network devices should compete or cooperate to get access to the channel, and the MAC protocols define how they should behave in case of the channel is available or busy. In general, those protocols follow a collision avoidance perspective, allowing broadcasts only when the channel is available.  Devices must refrain from transmitting while the medium is busy. Collision avoidance has been one of the most important challenges in wireless networks and its complexity has been intensified with the increasing number of devices and the requirements for Quality of Experience (QoE), mainly associated with real time applications, such as online video and voice transmissions.


Different works in the literature have indicated the low performance of Wi-Fi networks under dense scenarios. The elevated number of devices increases the probability of collisions. Also, the unplanned deployment of Wi-Fi networks arises in overlapping networks operating in a common frequency and interfering with each other. The popularization of Wi-Fi has caused the saturation of communication channels, mainly in large urban centers and public hotspots employing the 2.4GHz free license band.  Further, the requirements for QoE raise the complexity on access control mechanisms once they must ensure the quick transmission of urgent traffic without producing significant delay.

Hence, in order to cope with these new challenges, the IEEE created the Task Group AX to improve the 802.11 standard. They aim at increasing throughput and spectrum efficiency of highly dense wireless local area networks (WLANs). Their goal lies in offering a new amendment to 802.11 until 2019. Currently, a set of works has discussed models and evaluated methods, technologies and functional requirements to compose this new amendment. Also, researchers have proposed enhancements to the MAC protocols. Their goal lies in designing improved protocols considering the particularities of dense environments. 

In this context, this work contributes with the literature as follows. It strengthens the claims about the low Wi-Fi performance under dense scenarios by revisiting the collision problem. In particular, this article overviews the MAC protocols proposed for the 802.11 standard and discusses solutions to mitigate collisions. While previous publications about the 802.11ax offer a broader view~\cite{deng2014,bellalta2016}, our work studies the proposals focusing on collision avoidance. Finally, it contributes presenting future trends in MAC protocols, assisting in foreseeing expected improvements for the next generation of Wi-Fi devices.

This article proceeds as follows. First, we provide a background on current 802.11 standard and we revisit the problem of collisions related to MAC protocols. Then, we present an overview about the existing proposals for MAC improvements to 802.11ax amendment. Next, we discuss challenges, open issues and future directions in this field.

\section{Legacy IEEE 802.11 Wireless Networks}

During the last fifteen years, 802.11 compatible devices have proliferated in a worldwide scale. They have been largely deployed because of factors such as cost and simplicity. The 802.11 standard describes wireless communication structures for local range connectivity. It works with fixed or mobile devices. Since its first release in 1997, several amendments have been added to it. The Table~\ref{table:amendments} lists the main changes on each version. They have been introduced incrementally. The later versions keep the previous specifications to maintain backward compatibility.  For instance, the current version 802.11ac-2012 supports both 802.11n and 802.11a. And the next versions should continue supporting and coexisting with the current one.

\begin{table*}[!ht]
\centering
\caption{IEEE 802.11 Standard Amendments [adapted from \cite{deng2014}]}
\label{table:amendments}
\renewcommand{\arraystretch}{1.2}
\resizebox{\columnwidth}{!}{
\begin{tabular}{|l|c|c|l|}
\hline
Protocol  & Year & Band     & Main Changes                                                                         \\ \hline
802.11a  & 1999 & 5 GHz    & Orthogonal Frequency-Division Multiplexing (54 Mbps)                                 \\ \hline
802.11b  & 1999 & 2.4GHz   & High Rate Direct Sequence Spread Spectrum (11 Mbps)                                  \\ \hline
802.11c  & 2001 & N/A      & Incorporate bridging in wireless access points or bridges                           \\ \hline
802.11f  & 2003 & N/A      & Provides wireless access point communications                                        \\ \hline
802.11g  & 2003 & 2.4GHz & Extends throughput using the same 2.4 GHz band as 802.11b (54 Mbps)     \\ \hline
802.11i   & 2004 & N/A      & Improves authentication mechanisms to fix security flaws                         \\ \hline
802.11e  & 2005 & N/A      & Defines a set of Quality of Service (QoS) enhancements                               \\ \hline
802.11r  & 2008 & N/A      & Provides fast and secure handoffs from one base station to another                   \\ \hline
802.11n  & 2009 & 2.4/5GHz & Simultaneous transmissions for up to 4 clients by MIMO technology (150/600 Mbps) \\ \hline
802.11ac & 2012 & 5GHz     & Transmissions with up to 8 antennas and higher density modulations (until 6.77 Gbps) \\ \hline
\end{tabular}}
\end{table*}

\begin{figure}[ht]
\centering
\includegraphics[trim=30 220 20 90,clip, height=6cm]{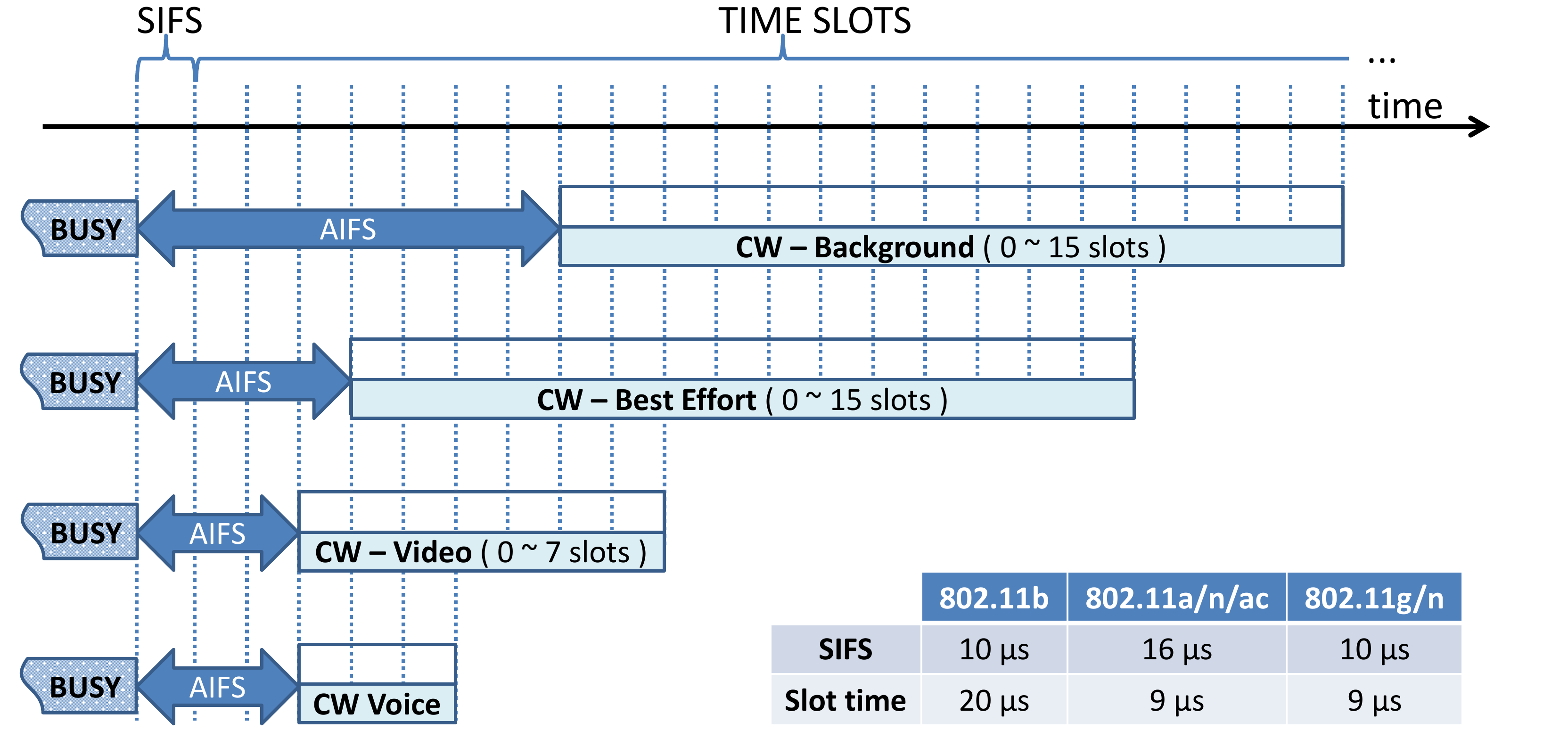}
\caption{Transmissions start at random slot within Contention Window for each traffic type.}
\label{fig:EDCA}
\vspace{-0.2cm}
\end{figure}

The MAC protocols used in all these versions are essentially the same. They follow a ``listen before talk'' approach in which a Carrier Sense is performed to check if the channel is clear (no other node is transmitting). If it is clear during certain periods of time, the nodes start their transmissions. These intervals between transmissions are important components in the coordination of the access mechanism. For instance, devices need to confirm every reception. Then, they send an acknowledgment (ACK) frame to notify transmitters about the communication successful. These ACKs must be sent as soon as possible after the reception. Thus, they send the ACK in the Shortest Inter Frame Space (SIFS). Further, only the receiver is allowed to transmit its ACK in this interval. And all other types of transmission must wait intervals longer than SIFS. This reduces the probability of collisions over ACKs. Other types of frames wait longer Arbitrary Inter Frame Spaces (AIFS). Accordingly to their priorities, devices define how long they must wait before starting their broadcasts. Finally, the mechanism adds a random time interval aiming to avoid simultaneous transmissions. This interval is inside of the Contention Window (CW). It works as a backoff countdown after some Inter Frame Space. The goal is to provide different access times among nodes with same priority level. The CW size varies depending on priority too as illustrated in Figure~\ref{fig:EDCA}. 

The random component is efficient to avoid collisions if there are few devices operating. It works because the CWs are divided into slots. Each device chooses one slot to start its transmission. If another device starts first, the Carrier Sense detects the busy channel. Then, the device interrupts its countdown, resuming in the next attempt. However, collisions would happen if two or more devices choose the same slot in their countdown. In this case, probably the signal would be unintelligible and the receiver would not send an ACK. When this occurs, the transmitters doubles their CW size and they try again. It reduces the chance of later collisions. But, they only observe a failure after the ACK timeout. Hence, this wastes the entire time transmitting and waiting for the ACK. Further, after a successful transmission, the CW size returns to its initial value. This causes instability and may provoke the capture of channel. Some devices can monopolize the channel if they can transmit while others have their CW expanded.  

\subsection{Collisions: a recurrent problem}

\begin{figure}[ht]
\centering
\includegraphics[trim=0 0 0 0,clip, height=8cm]{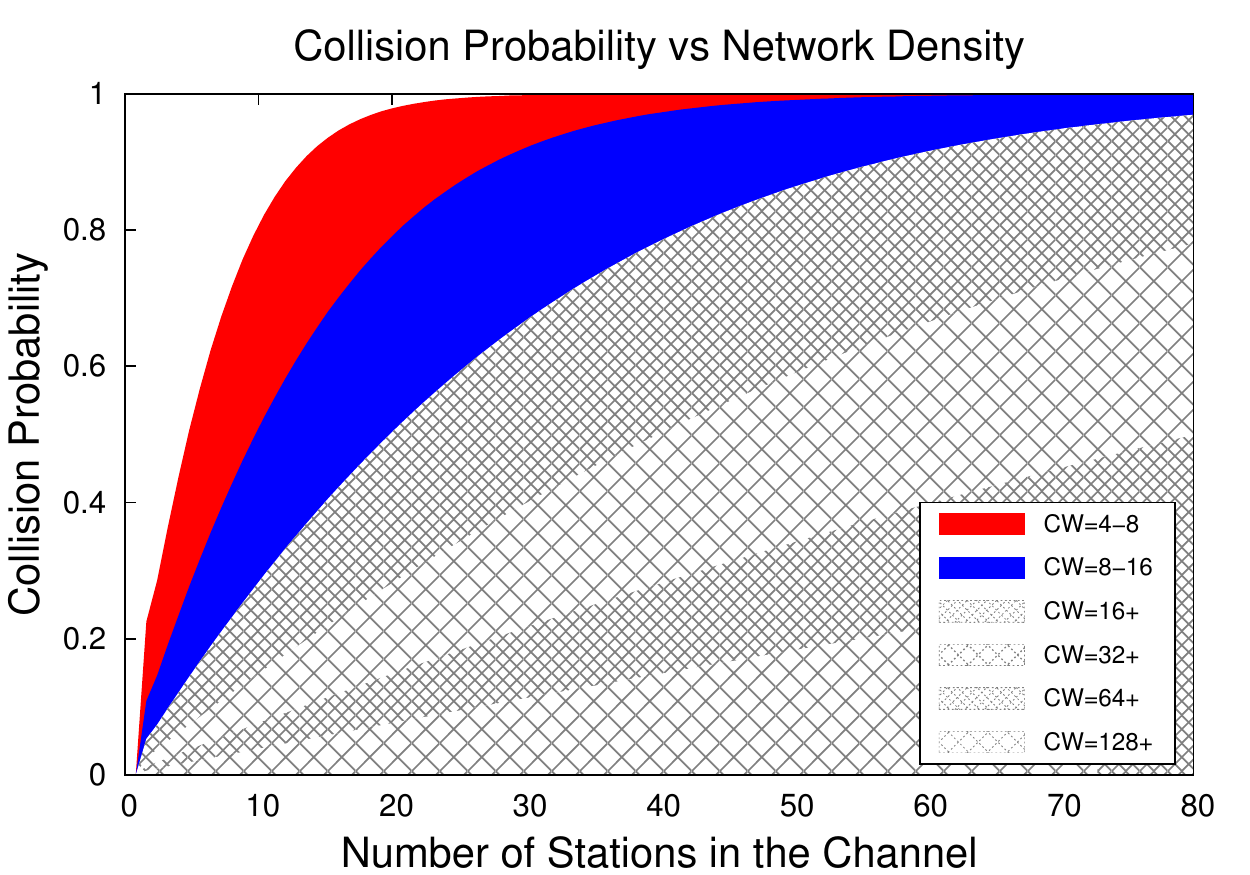}
\vspace{-0.2cm}
\caption{Collision Probability according CW size  (Voice=4$\sim$8 slots and Video=8$\sim$16 slots)}
\label{fig:ProbColl}
\vspace{-0.2cm}
\end{figure}

By combinatorial analysis, we calculate the collision probability according to the network density and CW size~\cite{dasilva} as illustrated in the Figure~\ref{fig:ProbColl}. The probability grows as network density increases, even more to smaller CW like that used for voice and video. From top to the bottom, the first and second tracks in the figure represent the collision probability for these types of traffic. The CW size for them is expanded only once until it reaches its maximum size. This results in a high chance of concurrent transmissions with just few nodes. For instance, the collision probability is higher than 50\% with just five nodes sending voice data. This analysis considers only collisions inherent to the mechanism, not taking into account anomalies caused by carrier sense fails or hidden nodes. The 802.11 MAC protocol uses a virtual carrier sense to prevent hidden terminal problem. Control frames are exchanged before data frames. They reserve the channel in the collision domain. But these frames are sent as well as conventional data frames. Hence, they suffer the same problems addressed above.

\section{MAC Protocols for 802.11ax: A Collision Avoidance Perspective}

The task group AX has worked on several proposals to improve the 802.11 standard. They have defined procedures for selecting and evaluating them. They also have specified deployment scenarios and parameters for testing. These scenarios contain a high number of stations and access points forming dense networks. Thereby, they emulate the problems of high collision rates. In this context, efficient MAC protocols are critical to improve the throughput and channel efficiency. The next sections provide an overview of proposed changes to them.  

\subsection{802.11 MAC Enhancements}

There are many studies about the current 802.11 MAC protocol. Mathematical analyses have demonstrated its behavior on performance, stability and fairness. Most studies indicate that larger windows are more suitable for dense networks, and smaller windows are more suitable for small networks. Thus, researchers came out to \textbf{Dynamic Adjustment of Contention Window}. Protocols following this approach can adjust the CW size according to the observed channel conditions. They avoid the access delay caused by a large initial window, as well as the high collision probability resulted from a small one. The challenge is how to define the ideal window size. In pursuit of this goal, researchers have proposed protocols based on medium use estimation. Measurements on Free Time or Channel Congestion can assist these estimations by each device. Exchanging this information among nodes may improve synchronization. Further, some protocols suggest a polynomial expansion of CW rather than an exponential one. Others suggest a smoother CW reduction after a successful transmission. There are examples of MAC protocols using these proposals such as DCWA, HBAB, EBA, AMOCW, DCWC, VBA and so forth. Until now, there is no consensus about which is better.

\begin{figure}[h!]
\centering
\includegraphics[trim=40 210 68 42,clip, height=4.5cm]{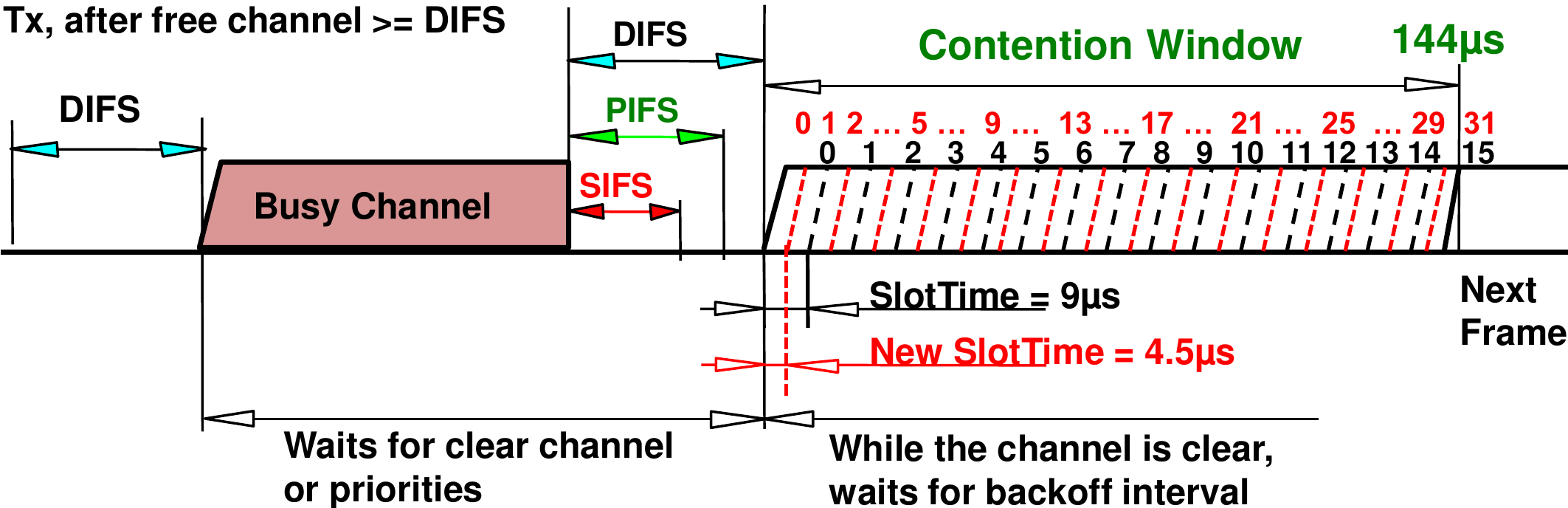}
\vspace{-0.1cm}
\caption{Proposed Changes to the Contention Window}
\label{fig:novoBackoff}
\end{figure}

Indeed, devices with large initial windows will be penalized if other devices keep their windows small. The proposal in~\cite{doc15-0914r1} suggests the \textbf{Enlargement of Initial CW}. The idea lies in increasing the number of contention slots while reducing the slot time. For instance, in OFDM, the slot would be reduced from 9$\mu$s to 4.5$\mu$s and the amount of slots would be doubled as illustrated in Figure~\ref{fig:novoBackoff}. Also, the CWs should be adjusted according to EDCA rules. They double their size in slots while maintaining the same total duration. Thus, new devices remain compatible with current standard if other intervals based on slot time are doubled too, e. g. AIFS and CWmin. 
 
The EICW is more relevant for voice and video traffic. It reduces the collision rates as exposed in our previous work~\cite{dasilva}. Further, the new devices would have more opportunities to send but without silencing the legacy. This should encourage upgrading. However, reducing slots may have impact on Clear Channel Assessment (CCA). Before transmitting, each device must check if the channel is not busy. With a reduced slot, they should do this on half time. Also, imperfections on CCA process can generate collisions. But, the recent hardware evolution leads to believe that it is possible to perform the CCA with enough accuracy even in such short time.

In another direction, some proposals suggest reservations based on Time Division Multiple Access. These \textbf{TDMA-like} protocols have a common CW size among all nodes. And each device reserves one slot to start its broadcasts. Usually, the first access is made choosing a random slot. If this transmission is successful, they defer from transmitting in all other slots. After a CW round they send the next frame in the same slot number. If a slot has been used for one device, no other can use it. This avoids collisions by reserving slots until they are liberated when not used. Protocols based in this idea are called \textbf{Deterministic Backoff}, e. g. ZC, L-BEB or CSMA/ECA and SRB. Usually they have problems with prioritization. But it can be solved allocating more than one slot per device. Unfortunately, this approach fails if neighbor devices do not cooperate. Legacy devices do not know about reserves. Hence, they could try to transmit in a reserved slot. This prevents the convergence to a state without collisions. 

Another way to reserve the channel access is allocating time intervals rather slots. As proposed in \textbf{TDuCSMA}~\cite{vesco2014} a service provider could assign a certain percentage of the bandwidth for each node or user. For instance, a Time Cycle (TC) of $100ms$ can be divided in Time Frames (TF) of $10ms$. Each TF is allocated for one station. During each allocated period, the selected node can transmit data without contending. It uses the shortest AIFS while the others continue applying the normal contention. The decision of access the channel is maintained in CSMA/CA. The mechanism is implemented on top of IEEE 802.11 architecture. This proposal can guarantee a Constant Bit Rate for demanding users. However, synchronization and coordination are open issues yet. Coordinators in neighboring networks need to avoid overlaps. And it is not clear how much the protocol depends on clock precision. This proposal was tested for audio, video and elastic data applications achieving good results.

\subsection{Discussion}

\begin{figure*}[!htb]
\vspace{-0.5cm}
\subfloat[][]{\includegraphics[trim=0 12 10 0,clip, width=.6\textwidth, height=7cm] {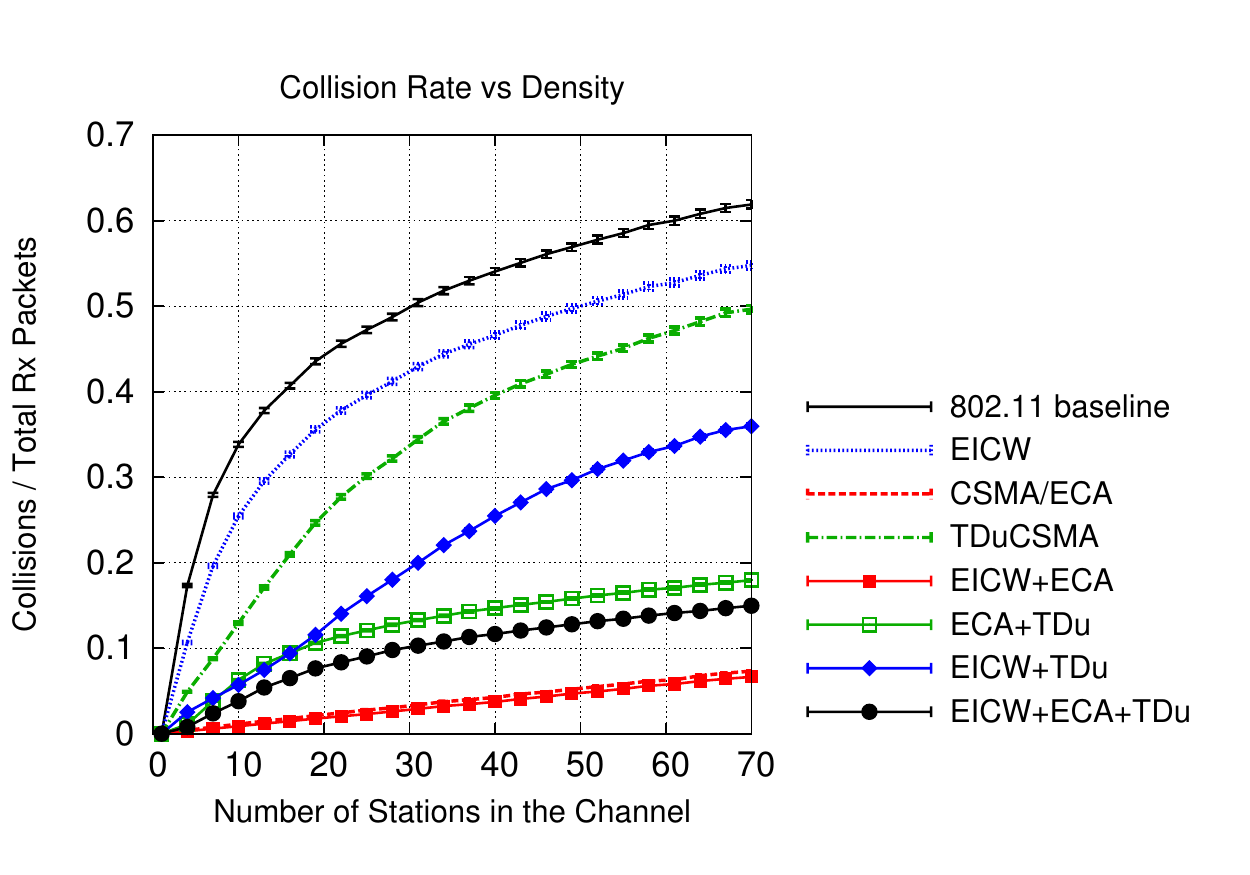}}\label{fig:label:1a}\hspace{-1.5cm}%
\subfloat[][]{\includegraphics[trim=0 0 10 0,clip, width=.55\textwidth, height=6.5cm] {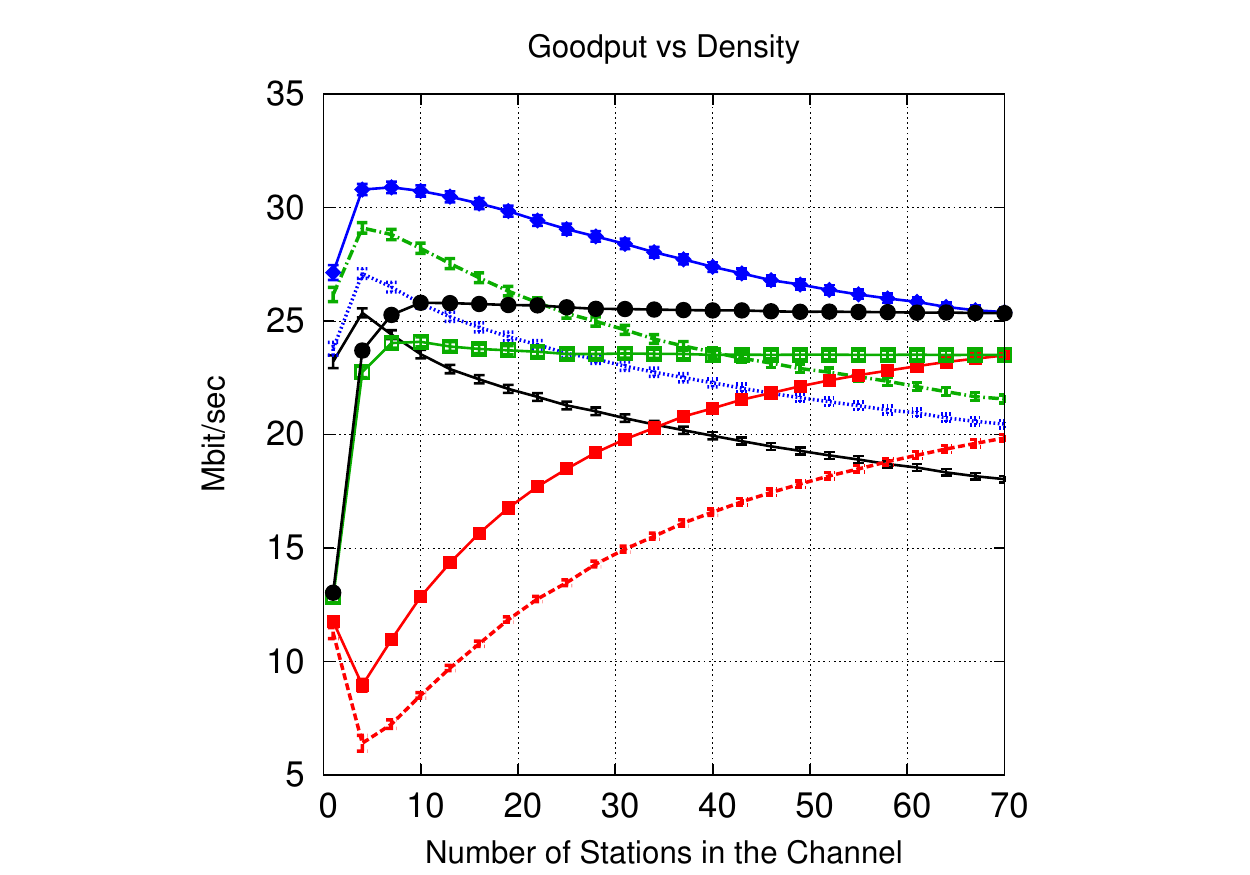}}\label{fig:label:2a}
\caption{Collision Rate and Goodput vs Network Density}
\label{fig:results}
\end{figure*}

We have analysed the main key enhancements proposed in the literature and reviewed in the previous subsection. Using the Network Simulator, version 3, we analyze \textit{collision rates per total of received packets} and \textit{throughput} comparing results from the following protocols: Extending the Initial Contention windows (EICW), time slots allocation without contention (TDuCSMA) and deterministic backoff (CSMA/ECA). In Figure~\ref{fig:results}, we observe that all protocols have successful reduced collisions in relation to the 802.11n standard baseline, over which 802.11ax is established. However, this reduction does not necessarily yields an increase in throughput. The simulations followed a scenario with stations sharing the same channel. They sent UDP data streams to their access points ensuing a generated traffic like the models which represent the typical predominant web traffic, but above the channel capacity. We have applied the same parameters used in the EICW evaluation in~\cite{dasilva} and we added the CSMA/ECA as available in~\cite{sanabria2016} and TDuCSMA as described in~\cite{vesco2014}. In each sample, we vary the number of stations to measure the performance of each protocol as the scenario becomes denser. Further, we also combine the protocols applying them in pairs and together. 


For the Collision Rate, CSMA/ECA achieves the best results individually. The reduction in collision is a little bit better if ECA is combined with EICW. But TDuCSMA degrades performance when combined with ECA. TDu changes the determined slot defined by ECA on every time frame, forcing a new convergence process. In the other hand, when EICW and TDu are simultaneously applied, they amplify their collision reduction. Intuitively, we can assume that fewer collisions will result in a better throughput caused by less waste in channel time. However, the reduction in collisions observed in ECA has no corresponding in throughput increasing. The problem is the timeout in the queues. While waiting their turn on ECA, the packages reach the limit time delaying before an opportunity to transmit. This shows that a collision-free protocol does not always deliver the best performance.

Although the observed results may vary according to the volume and type of traffic, we notice that the EICW combined with TDuCSMA have gained the highest throughput in major evaluated scenarios, including triple play. But, researches on these proposals are incipient in the context of TGax. There is a lack of studies or benchmarks comparing them. This is an opportunity for future researches. Probably, experiments on these MAC enhancements will anticipate the next standards in channel access.

\section{Multi User MAC Protocols for 802.11ax: A New Challenge}

TGax group has discussed other improvements like spatial efficiency and interference mitigation in order to reduce collisions and improve performance. Since 802.11n, MIMO has been used to send many streams at same frequency band. This method divides the channel into parallel subchannels. While each subchannel has lower speed, the parallelism results in higher aggregated speed. MIMO-OFDM uses multiple antennas turning their signals through right-angles (orthogonal). Hence, they can be easily decoded even stacked closer together. At first, MIMO has been employed to increase the throughput between one transmitter and one receiver. Since 802.11ac, it has been employed to provide simultaneous transmission to many receivers. In this case, every subchannel carries data for one receiver but only one can transmit them at time. Thus, the channel access method is the same as a single access. Nevertheless, 802.11ax probably adds multiple transmitters in MIMO. This demands changes in MAC protocols to support multiple accesses. They should allow more than one device accessing the channel, but each one in its subchannel.

\begin{figure}[ht!]
\centering
\includegraphics[trim=40 315 100 80,clip, height=4.0cm]{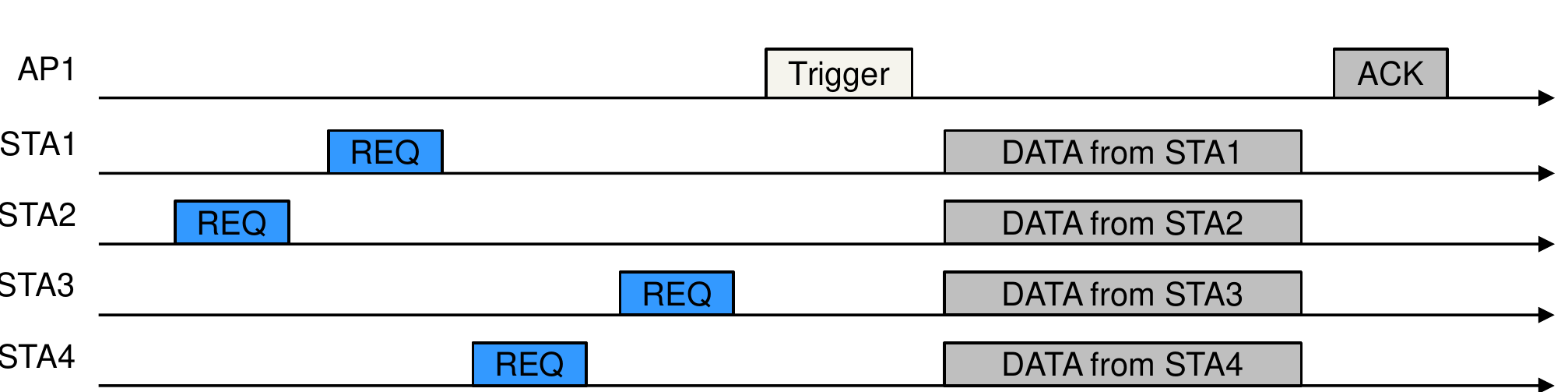}
\vspace{-0.5cm}
\caption{Uplink Multi User MIMO Trigger Process}
\label{fig:TriggerFrame}
\end{figure}

In the literature, there are dozens of proposals on multi-user MAC protocols. TGax should adopt a scheduled coordinated method. Coordinated means devices should exchange control frames to reserve the channel. And scheduled means the AP should trigger the concurrent transmissions. 
To reduce overhead, the requests for an uplink may be inserted into a normal frame. The stations send their common frames just adding one bit into frame header. It informs they have more data to be sent \cite{doc15-0608r1}. Then, the AP groups these stations and triggers the simultaneous access. It sends a control frame setting which stations can transmit. This frame must contain information about frequency offset, spatial stream, tone, length and so on. This allows the stations to start their transmissions in a synchronized way as illustrated in Figure~\ref{fig:TriggerFrame}. Once the control frame reserves the channel, the other devices are silenced avoiding collisions. However, the AP access procedure is not clear yet. The trigger frame could be sent after a SIFS or follow EDCA rules or other method as PCF. 

Another promising technique for multi user transmissions is OFDMA. It greatly improved performance of LTE and WiMax networks where it is already in use. OFDMA assigns different OFDM subchannels for each user. By that, transmitters can use their assigned subchannels to transmit without collide with others. But, in this case, only the carrier locking does not imply that transmissions are not allowed \cite{doc14-1428r0}. It makes difficult to check if the channel is clear or busy in CCA. This is another reason in favor of AP-coordinated MAC protocols. Since 802.11n, the APs have to get Channel State Information (CSI) from stations for downlink MIMO. This provides a better view of their domain. The current CSI feedback may be improved to add more data about PHY layer, buffers and collisions. The coordinators could make better decisions with this information. For instance, if APs know the type of collisions then they could distinguish between normal or anomalous collisions. They could apply the appropriate protection if collisions are caused by hidden terminals. 
The challenge is provide CSI feedback without introduce much overhead.

\subsection{Full-Duplex MAC Protocols}

Although simultaneous transmissions may cause interference, they can be used to increase the channel efficiency when implemented in a controlled way. Signal Interference Canceling (SIC) techniques have been used to provide In-Band Full-Duplex communication. Usually, devices cannot receive while transmitting in the same channel. It happens because their own strong signal interferes on other signals from distant sources. But, SIC removes this interference taking advantage of devices know their own signal. It results in their peer's signals and enables communication. And this method does not have costs with extra band to provide FD. Evaluation results have shown until 300\% of throughput enhancement in TGax proposed scenarios~\cite{7041186}.

But, IBFD introduces new challenges to MAC protocols. For instance, other devices detect a collision when two near devices use IBFD. Thus, they apply Extended Inter Frame Spaces (EIFS) much longer than normal IFS. Further, the current RTS-CTS control frames cannot properly reserve the channel for IBFD. The frames could have different lengths. Then the device sending the RTS cannot stand the other frame length. And the frames are not transmitted exactly at same time. Hence, the mechanism must be adjusted to (a) exchange ACK frames, (b) solve the hidden node problem in asymmetric data traffic and (c) avoid starve other nodes. However, discussions about IBFD are still incipient in TGax forums. More details about the MAC issues for IBFD systems can be found on the survey in \cite{7041186}.

\subsection{Other improvements}

The recent 802.11ac amendment provides MSDU and MPDU aggregation. Frame aggregation decreases MAC header, IFS and other extra overhead. Further, the contention time is incurred only once in the transmission of the large frame instead of many small packets. But, the probability of successful transmission is higher for smaller frames due wireless errors. An Adaptive Block Frame Length is discussed to set the ideal length \cite{doc14-0613}. In another line, some proposals suggest adding some management functions to the AP. They could determine priorities based on STA characteristics giving more opportunities for faster devices~\cite{doc13-1349r0}. Or they could cooperate. An Access Controller can coordinate them in managed deployed networks, e. g. campus or conference environments \cite{doc13-1157r3}. Wifox, ADWISER and MegaMIMO are practical examples of dense networks operating with a control entity. Further, APs can share Beacon frames among neighbour BSS. This could reduce network overhead with control frames. And beacon frames could be prioritized to avoid collisions.

Dynamic Sensitivity Controls have been discussed to achieve CSMA/CA optimizations \cite{doc16-0310}. They regulate the relative size of contention domains which affects throughput. Simulation results have shown gains in managed networks as well in dense residential apartment buildings. Adaptive configuration of CCA threshold (or receiver sensitivity) can reduce the network operating range. This decreases the collision domain area. But DSC are dependent to scenario setup and path loss parameters. Collisions increase when using higher CCA levels. And spatial reuse is suppressed when CCA level is lower. Thus, MAC Protocol should adapt the backoff procedure according CCA level. For instance, the backoff countdown would continue even if channel is busy since the signal is weak enough to interfere.  
Beyond this, the backoff process can be migrated to signaling through OFDM subcarriers \cite{doc14-0589r0}. Back2f and WiFi-BA protocols are examples on this type of access mechanism. This would reduce the contention time and avoid collisions. But these proposals require additional hardware such as SIC structures used in IBFD. However, they could emerge for 802.11ax if they afford backward compatibility and also if their costs are feasible. 

\section{Conclusion}

High efficient wireless networks have been a topic of interest to researchers from academia and industry. Since it was constituted, TGax group works on improve throughput and spectrum efficiency in 802.11 standard. They have discussed several proposals to solve network problems in dense environments. Although there is still a long time for its completion, the amendment drafts suggest some improvements. This article gave an overview of these improvements related to Medium Access Control protocols in particular giving a focus on collision avoidance. This work also contributed by revisiting the collision problem considering now highly dense scenarios. Collisions have become increasingly critical with the rising number of users. Further, this article also gave trends leading to the near future implementation. The current themes are classified and discussed. There are many proposals on the agenda. 
However, this article contributed to a basis for foreseeing the next generation Wi-Fi networks.


\bibliographystyle{IEEEtran}
\bibliography{IEEEabrv,scibib}





\end{document}